# DNA sequence-dependent mechanics and protein-assisted bending in repressor-mediated loop formation

James Q. Boedicker[1], Hernan G. Garcia[2], Stephanie Johnson[3], Rob Phillips[1]

[1]Departments of Applied Physics and Biology, [2]Department of Physics, and [3]Department of Biochemistry and Molecular Biophysics, California Institute of Technology, 1200 California Boulevard, Pasadena, CA 91125.

* Corresponding author phillips@pboc.caltech.edu.

**ABSTRACT**

As the chief informational molecule of life, DNA is subject to extensive physical manipulations. The energy required to deform double-helical DNA depends on sequence, and this mechanical code of DNA influences gene regulation, such as through nucleosome positioning. Here we examine the sequence-dependent flexibility of DNA in bacterial transcription factor-mediated looping, a context for which the role of sequence remains poorly understood. Using a suite of synthetic constructs repressed by the Lac repressor and two well-known sequences that show large flexibility differences *in vitro*, we make precise statistical mechanical predictions as to how DNA sequence influences loop formation and test these predictions using *in vivo* transcription and *in vitro* single-molecule assays. Surprisingly, sequence-dependent flexibility does not affect *in vivo* gene regulation. By theoretically and experimentally quantifying the relative contributions of sequence and the DNA-bending protein HU to DNA mechanical properties, we reveal that bending by HU dominates DNA mechanics and masks intrinsic sequence-dependent flexibility. Such a quantitative understanding of how mechanical regulatory information is encoded in the genome will be a key step towards a predictive understanding of gene regulation at single-base pair resolution.

**INTRODUCTION**

The regulation of the genetic information in bacteria and eukaryotes alike often involves the physical deformation of the DNA molecules that carry this information.  DNA deformations enable interactions between distant DNA-bound proteins by bringing them into close proximity [1], regulate the accessibility of DNA binding sites in eukaryotes by sequestering them in nucleosomes [2], and even can modulate binding affinities of transcription factors [3].  Indeed, long range interactions in the eukaryotic setting are the rule rather than the exception [4], as is regulation by the binding of multiple transcription factors, which in many regulatory architectures necessitates loop formation [1].  In addition, many classic bacterial transcription factors such as Lac repressor [5-8], $\lambda$ repressor [9], Gal repressor [10, 11], NtrC [12-14], and AraC [15] all involve loop formation as an intrinsic part of their regulatory mechanism [16].  The deformation of DNA is critical to regulation.  Here our aim is to quantitatively understand the role of DNA mechanics in promoter response using a combination of *in vivo* and *in vitro* experiments and predictive models of gene expression.

The majority of our knowledge of DNA mechanics comes from a gamut of *in vitro* experiments that include techniques as diverse as cyclization and molecular scale imaging techniques such as AFM and electron microscopy [17].  Such *in vitro* experiments have established that double-stranded DNA is highly dynamic even at short length scales and that the DNA sequence influences its bending rigidity and torsional modulus [17-23, 24 ], leading to speculation about the role of DNA mechanics in gene regulation *in vivo*.

The role of the *in vivo* deformability of DNA in gene regulation is already a source of debate and controversy in the context of nucleosome formation [24-30].  One of the hypotheses that has been advanced for the physical mechanism of nucleosome positioning highlights intrinsic, sequence-dependent flexibility of DNA as a key factor [21, 23, 25, 30-33].  Several natural as well as synthetic sequences have been identified that strongly favor nucleosome formation [31], a propensity which has been shown to directly correlate with DNA flexibility in the absence of any histones [24], as measured by *in vitro* ligase-mediated cyclization assays, although subsequent experiments have brought these results into question and alternative hypotheses have been proposed for the role of sequence in nucleosome positioning [29, 34].  Despite the controversy, these experiments suggest that regulatory decisions in cells might be influenced by the mechanical properties encoded in the DNA sequence in addition to those properties associated with protein coding regions and protein binding sites.  In the eukaryotic setting, this can take the form of a competition between occupancy by nucleosomes and transcription factors [2, 35, 36].  Although nucleosome positioning is an important part of regulatory decisions in eukaryotes, DNA mechanics likely has a much broader role in gene regulation, even in bacteria where nucleosomes are absent.  Here we examine whether sequence-dependent mechanical effects are evident in another context, namely loop formation in bacterial transcriptional regulation, as shown in Figure 1.

Bacterial regulatory architectures that feature upstream activators, such as the activator NtrC in *E. coli* or analogous *nif* promoters of *Klebsiella pneumoniae*, have been used as model systems to study the role of DNA mechanics and bent sequences of DNA in loops that form between upstream activators and $\sigma^{54}$ RNA polymerase [12, 14, 37, 38]. It has been shown that transcription from this complex requires DNA bending, either assisted by proteins such as IHF [38] or by sequences known as poly-A tracts, which adopt a statically bent conformation [37]. It should be noted that although bending of the poly-A tracts assists in forming contacts between RNA polymerase and upstream activators, poly-A tracts can also influence expression through direct interaction with the α subunit of RNA polymerase [39]. Poly-A tracts have also been shown to stabilize Lac repressor-mediated loops *in vitro* [40]. These earlier studies have focused on sequences that adopt intrinsically *bent* conformations, but have not been extended to intrinsically *bendable* sequences which have more isotropic flexibility [41]. Here we build upon those experiments by extending the breadth of sequences and regulatory architectures in the context of an underlying predictive theoretical framework. We characterize the specific quantitative details of the role of DNA mechanical properties on the regulatory input-output function by comparing two sequences that have been shown to differ in their intrinsic bendability, without adopting specific, statically bent conformations [30], and quantify the ability of these sequences to form Lac repressor-mediated loops.

Many previous studies have made use of the *lac* operon to study a variety of aspects of gene regulation both with and without loop formation [5-7, 42]. The *lac* operon is negatively regulated by Lac repressor (LacI), and loops DNA as part of its regulatory mechanism. Looping increases repression up to 50-fold with respect to regulation by the main operator alone [7, 42]. There is a rich history of applying theoretical models to understand the contribution of looping to gene regulation by Lac repressor [6, 43-48]. Thermodynamic models can be used as an intellectual prism not only for interpreting gene expression measurements and extracting key microscopic parameters such as the looping free energy, but more importantly, for making quantitative, testable predictions about the regulatory variables [47, 48]. Predictive biophysical models have provided a useful conceptual framework for dissecting how tunable parameters such as transcription factor numbers, binding strengths, and binding site locations influence gene regulation [47, 49, 50]. It is this kind of predictive framework that interests us here. In order to quantitatively understand sequence-dependent mechanical properties as a control parameter for gene regulation, we implement such a theoretical framework to make precise predictions as to how sequence modulates loop formation and loop-mediated repression. We then test those predictions using *in vivo* transcription assays and *in vitro* single-molecule assays.

Our thermodynamic model predicts a dramatic dependence of the level of gene expression on the flexibility of the looping sequence. To test this prediction, we incorporated a sequence of DNA shown to be highly flexible from *in vitro* cyclization assays into a Lac repressor-mediated loop. It was known from previous results in single-molecule *in vitro* experiments that the flexible sequence more readily formed DNA loops as compared to a random sequence [51]. Surprisingly, the sequence that more easily forms loops *in vitro* did not lead to increased repression *in vivo* in a wild-type host. We hypothesized that this lack of sequence dependence *in vivo* could be due to the activity of protein factors which modulate chromosome structure, as it has been well established that DNA-structuring proteins such as HU and IHF assist in restructuring the DNA to accommodate regulatory interactions [11, 38, 52], including loop formation by the Lac repressor [44]. We determined that these proteins also buffer the influence of intrinsic sequence flexibility in loop formation *in vitro*, and show that deletion of the nucleoid-associated protein HU restored sequence dependence to DNA looping *in vivo*. Through the combination of predictive biophysical models and quantitative experimental measurements, we identified a previously unknown suppression of sequence-dependent elasticity in the context of gene regulation. These results expand our understanding of the DNA mechanical code embedded within the genome, and emphasize that a predictive description of DNA looping combining an analysis of the interplay of transcription factors, DNA sequence and nucleoid-associated proteins can lead to the discovery of new regulatory mechanisms.

**MATERIALS AND METHODS**

**Strain construction**

Site-directed mutagenesis was used to move the auxiliary operator further upstream in one base pair increments as the looping sequence was incorporated between the auxiliary operator and the promoter. Strains containing deletions of *hupA* and *hupB* were obtained from the Keio collection. These deletions were transferred into host strains by P1 phage transduction, see Supplementary Figure 14. The random sequence E8 and the flexible sequence TA were based upon sequences used [30]. See Supplementary Information for more details and Supplementary Tables 3 and 4 for the sequences used.

**Gene expression measurements**

The inhibition of gene expression due to repressor was quantified by comparing the ratio of gene expression in cells with and without repressor, as in Equation 1. Cultures of cells were grown to exponential phase in M9 media containing 0.5% glucose and measured using a plate reader. Cells not containing the fluorescent reporter were also grown to determine the autofluorescent background. Measurements were normalized by dividing by the optical density of each culture at 600 nm. For handling of measurement statistics see [48].

**Single-molecule measurements**

Tethered particle motion experiments were performed exactly as described in [51, 53]. Briefly, PCR with labeled primers was used to create double-stranded, linear DNA constructs for TPM from the promoter regions of the E8- and TA-containing constructs used in the *in vivo* experiments. Each TPM construct contained 56 to 89 bp of either the E8 or TA sequence and 36 bp of the promoter. The looping sequence was flanked by the Oid and O2 operators and about 150 bp of DNA between Oid and the coverslip or O2 and the bead. The looping probabilities for these constructs were measured in the presence of 100 pM Lac repressor purified in-house.

**HU purification**

Heterodimeric HU was purified from strain RJ5814 (*ihfB::cat fis::kan-767 endA*::Tn*10 his ilv $\lambda$cI857 N$^+$*, containing plasmid pP$_L$-*hupAB* from Roger McMacken, [54]), a kind gift from Reid Johnson, according to a protocol modified from [55]. Details are given in the Supplementary Information.

**RESULTS AND DISCUSSION**

**Thermodynamic model of loop-mediated gene regulation**

Thermodynamic models have been widely used as a quantitative framework to describe transcriptional regulation in bacteria [56-58], including in the analysis of gene regulation involving DNA looping [6, 43-48, 59]. We have previously used a particular class of such models to explore, from both a theoretical and an experimental perspective, how each parameter of these models is a "knob" modulating gene regulation in the case of simple repression [57]. More recently, we used these thermodynamic models to validate how the knobs of operator binding energies and number of repressors per cell tune repression in the more complicated case of repression by loop formation [48]. In this paper, we examine the role of the looping sequence in gene regulation, an important remaining knob for this model system which has largely been overlooked in previous studies. Using the tools of statistical mechanics, we can predict the values of various experimentally defined quantities, such as changes in gene expression levels. In particular, we are interested in changes to *repression* of gene expression, which we define as the gene expression in a strain that lacks the Lac repressor, relative to the gene expression in a strain that harbors the Lac repressor, as given by

$$\text{Repression} = \frac{\text{gene expression}(R=0)}{\text{gene expression}(R \neq 0)}. \quad (1)$$

In order to derive a model of repression that accounts for DNA looping by the Lac repressor, we first enumerate the different states of the system (for example, an RNA polymerase (RNAP) bound to the promoter, or a repressor bound to an operator), and assign corresponding statistical weights and relative rates of transcription for each state. The states and weights for our particular model are diagrammed in Figure 2A.

In Figure 2A, $P$ is the number of RNAP molecules per cell, $\Delta\varepsilon_{rmd}$ is the binding energy of Lac repressor to the main operator, $\Delta\varepsilon_{rad}$ is the binding energy of Lac repressor to the auxiliary operator, $\Delta\varepsilon_{pd}$ is the binding energy of RNAP to the promoter, $N_{NS}$ accounts for the number of nonspecific binding sites for repressor, $\beta$ is the inverse of Boltzmann's constant times temperature, $R$ is the number of Lac repressor molecules per cell, and $\Delta F_{loop}(L)$ is the free energy needed to form a loop of length $L$ bp. $\Delta F_{loop}(L)$ is free energy change of the looped complex minus the free energy changes of the Lac repressor binding to the two operators independently. We note that here $R$ is the number of Lac repressor tetramers, and is therefore multiplied by 2 to account for each tetramer having two binding heads (each dimer can bind to a single operator). The values of all parameters with the exception of $\Delta F_{loop}(L)$ are known from previous experiments and listed in Supplementary Table 5. "Rate" denotes the rate of transcription initiation from of each particular state.

These thermodynamic models are based upon the assumption that the probability of finding the system in a given regulatory state is a function of the free energy associated with each state of the system. If the binding and unbinding of RNA polymerase to the promoter and Lac repressor to the operators are in quasi-equilibrium with respect to the rate of transcription initiation, then the Boltzmann distribution tells us that the probability of a given state $i$ is equal to

$$p(state_i) = \frac{weight_i}{\sum_{i=1}^{n} weight_i} = \frac{weight_i}{Z}, \quad (2)$$

where the weight is related to the energy of that particular state as shown in Figure 2A. The denominator of Equation 2, the partition function often written as $Z$, is the sum of the weights of all states. In this context the rate of transcription can be obtained by calculating the probability of being in each state that allows transcription, multiplying them by their respective transcription rate and adding them all up. For example for the promoter architecture in Figure 2A we have

$$\text{gene expression} = k_2 p(state_2) + k_3 p(state_3) = \frac{k_2 \frac{P}{N_{NS}} e^{-\beta \Delta \varepsilon_{pd}} + k_3 \frac{P}{N_{NS}} \frac{2R}{N_{NS}} e^{-\beta(\Delta \varepsilon_{pd} + \Delta \varepsilon_{rad})}}{Z}, \quad (3)$$

in which $k_2$ and $k_3$ are the rate constants for transcription initiation from states 2 and 3.

Given the definition of repression found in Equation 1 and the probability of gene expression in Equation 3, we derive an expression for loop-mediated repression. A few simplifying approximations detailed in the Supplementary Information result in,

$$\text{Repression}_{loop} = \frac{1 + \frac{2R}{N_{NS}}(e^{-\beta \Delta \varepsilon_{rad}} + e^{-\beta \Delta \varepsilon_{rmd}}) + \frac{4R(R-1)}{(N_{NS})^2} e^{-\beta(\Delta \varepsilon_{rad} + \Delta \varepsilon_{rmd})} + \frac{2R}{N_{NS}} e^{-\beta(\Delta \varepsilon_{rad} + \Delta \varepsilon_{rmd} + \Delta F_{loop}(L))}}{1 + \frac{2R}{N_{NS}} e^{-\beta \Delta \varepsilon_{rad}}}. \quad (4)$$

Equation 4 gives us a quantitative input-output function for constructs such as the one schematized in Figure 1A. As noted above, we have recently demonstrated the validity of the model leading to Equation 4 by systematically testing some of the regulatory system's tunable "knobs" depicted in Figure 2B. Previous work has focused on the parameters of protein copy numbers and binding energies [48]. Here we view these parameters in Equation 4 as fixed and explore a key remaining model parameter, namely the energy associated with forming a loop in the DNA, $\Delta F_{loop}(L)$.

### Thermodynamic model predicts a dramatic effect of sequence flexibility on repression by DNA looping

A strategy for modulating the looping free energy is to exploit the sequence-dependent mechanical properties of DNA. To determine how changes in the bendability of the DNA in the loop alter the level of gene expression through loop formation, we expanded a library of synthetic constructs where the length and sequence of the DNA looped by Lac repressor is varied systematically [5, 6, 42, 48]. In an earlier work we reported that loop-mediated gene expression was modulated by key regulatory parameters such as protein copy numbers and binding energies [48]. In this previous work, looping sequences were based on the E8-94 sequence described by Cloutier and Widom [24], which we call the 'random' sequence. Here we characterized new constructs containing the nucleosomal positioning sequence 601TA (henceforth called TA or the 'flexible' sequence). TA is much more likely to form a DNA minicircle than the random sequence E8 [24, 30]. The sequences we use here, E8 and TA, represent the extremes of flexibility found in previous studies [24, 30], as shown in Figure 2C. Therefore although we only measure two sequences, they encompass a broad range of sequence-dependent flexibilities. From these previous measurements on the propensity of these DNAs to form *circles* and *nucleosomes*, we calculated that the flexible DNA sequence has a looping energy lower than the random sequence by as much as 2-3 $k_BT$, depending on the length, as shown in Supplementary Figure 1. As a result, we expect the flexible sequence to lead to more looping, and therefore more repression, than the random looping sequence.

Starting with our previously reported data for the random looping sequence [48], we predict how repression will change when the looping region is replaced with the flexible looping sequence. Figure 3A shows the repression as a function of length of the random sequence. The approximately 11 bp period present in the curve is associated with the twist of the DNA as the binding sites are moved with respect to each other [42]. It is interesting to note that this energetic cost to twisting at lengths when the operators are not in alignment is reduced as the loop length increases. This figure also highlights the rather counterintuitive result that DNA easily forms loops shorter than its persistence length of 150 bp. We use Equation 4 to extract the *in vivo* mechanical properties of DNA through $\Delta F_{loop}(L)$ from these measurements of repression, as shown in Figure 3B for the random looping sequence. Extracting $\Delta F_{loop}(L)$ from experimental data using the model in Equation 4 does not involve any unknown parameters since all other model inputs (the binding energies, average number of repressors per cell, and the size of the *E. coli* genome) are known and are summarized in Supplementary Table 5. Equation 4 also allows us to quantitatively predict the change in repression that we should expect due to changes in the flexibility of the sequence in the loop.

In Figure 3C we plot the expected change in repression as the looping energy is decreased by tuning the flexibility of the looping sequence. The vertical axis shows the relative change in repression as the looping energy is decreased by 0 to 2 $k_BT$ with respect to initial looping energies between 0 and 20 $k_BT$. The red shaded region represents the approximate range of looping energies measured in Figure 3B, and indicates that gene regulation in our constructs will be sensitive even to small changes in the looping energy. It is interesting that when $\Delta F_{loop}(L)$ is more than 14 $k_BT$ the extent of repression is not sensitive to a reduction in the looping energy up to 2 $k_BT$. Loop formation at these looping energies is energetically costly and becomes a low probability state which does not significantly contribute to repression, as shown in Supplementary Figure 2.

Figure 3D shows the prediction of how repression will change for the constructs measured in 3B if the looping energy is reduced by 1 or 2 $k_BT$. Even a 1 $k_BT$ change in the looping energy increases repression more than 2 fold. Given previous *in vitro* cyclization measurements on the sequences E8 and TA showing more than a 2 $k_B$T difference in their looping energies (Supplementary Figure 1), replacing the random sequence in the loop with a flexible sequence should greatly increase repression *in vivo*.

**Loop formation with Lac repressor is sequence-dependent *in vitro***

The relationship between sequence and flexibility in cyclization versus looping is not always straightforward, with flexibility in one assay not always translating to flexibility in another. Boundary conditions involved in forming a protein-mediated DNA loop are different than those for a ligated DNA minicircle (see [43, 60-62] for more details on these subtleties). As a result, before examining the effect of the different looping sequences on gene expression *in vivo*, we used previous *in vitro* results to calculate the sequence-dependent change in the looping free energy [51].

To verify that these two sequences do in fact behave differently from each other in the context of loop formation, as they do in cyclization assays, we turned to a single-molecule assay called tethered particle motion (TPM). The TPM assay observes the formation and breakdown of Lac repressor-mediated loops directly, in the absence of complicating factors in the cell [51, 63-65]. The change in the length of the linear double-stranded DNA tether attached to a microscopic bead, as shown in Figure 4A, quantifies the probability of loop formation in the presence of Lac repressor. Previously we have reported TMP measurements of the probability of looping for several of the random and flexible looping sequences used in the *in vivo* experiments described here [51]. Here we reproduce these data in Figure 4B to demonstrate to what extent sequence influences *in vitro* looping.

The putative flexible constructs have a higher looping probability than the random constructs *in vitro* (Figure 4B). Here we apply a similar statistical mechanical model to that described above for *in vivo* repression to extract the *in vitro* free energy of loop formation [51, 53]. The resulting *in vitro* looping free energies are shown in Figure 4C. Converting these previously reported looping probabilities to free energies of loop formation helps clarify whether the sequence-dependence of looping measured *in vitro* is sufficient to result in an observable change repression. The energies of the flexible versus random sequences differ by 1 to 3 $k_B$T at most operator distances. This shows unequivocally that the process of loop formation itself, at least outside the cell, depends on flexibility of the DNA sequence. Moreover, as shown in Figure 3D, we expect that such differences in the looping free energies of 1 $k_B$T or more should be easily detectable as changes in repression levels *in vivo*.

***In vivo* repression is not dependent on the sequence of the DNA in the loop**

To quantify how the flexibility of the looping sequence influenced gene regulation, we transferred the *in vitro* DNA constructs into the genome of *E. coli* and measured regulation of YFP. These results are shown in Figure 5A. Surprisingly, and in sharp contrast to both *in vitro* cyclization and looping results, the putatively more flexible sequence of DNA in the loop leads to no measurable increase in repression *in vivo*, entirely at odds with the prediction shown in Figure 3D. Figure 5B shows an overlay of the looping free energies for both sequences, reiterating that there is no difference in loop formation of the two sequences considered here. Although it is possible that other looping sequences not tested would show a difference in repression, recall from Figure 2C that the two sequences compared here correspond to a large range of *in vitro* sequence-dependent flexibility as measured by DNA cyclization [24, 30]. Furthermore, *in vitro* looping experiments show that the rank ordering of the sequences is maintained with respect to cyclization, as the flexible sequence has a lower energy than the random one for all lengths considered (Figure 4C). Thus, these two sequences represent the extreme case to test sequence-dependent mechanical effects.

The contrast between the *in vivo* and *in vitro* behavior is more clearly revealed by plotting the difference in looping energy between the random and flexible sequences in each context (Figure 5C). For the *in vivo* data, the looping energy differences are scattered with most values close to zero within experimental error. On the other hand, *in vitro* the differences in looping energy are greater than zero for almost all lengths tested.

Given such strong sequence dependence to loop formation *in vitro*, why is there no difference in repression *in vivo* using the exact same DNA constructs? One of the key differences between the *in vivo* and *in vitro* settings is the state of the DNA: *in vivo*, the genome is supercoiled and is decorated with both specific and nonspecific DNA-bending proteins. It has been shown previously that some of these nonspecific DNA-bending proteins can alter the mechanical properties of DNA both *in vitro* [14, 52, 66, 67] and *in vivo* [11, 38]. For example, the DNA-bending protein HU is required for GalR looping [11], and IHF is required for some NtrC-like loops [38]. Random, nonspecific binding of HU to DNA results in many flexible bends in the genome which accommodate a large range of bend angles [68]. Although the Lac repressor can form loops in the absence of any additional factors, it has been shown that HU increases Lac repressor-mediated loop formation *in vivo* [5]. However, the interplay between the intrinsic flexibility of the DNA set by the sequence and these DNA-structuring proteins is not understood. That is, it is not clear whether these DNA-bending proteins increase looping by all sequences equally or if they allow loop formation through a mechanism that overcomes the intrinsic flexibility or inflexibility of particular sequences.

**Deletion of the DNA-bending protein HU restores sequence-dependent loop formation *in vivo***

Constructs containing loops with the random or flexible sequences were integrated into a host strain lacking the DNA-bending protein HU. Similar strains have been used previously to quantify the role of HU in loop formation [5, 44, 69]. We asked whether in addition to lowering the free energy change of loop formation, as previously shown in [5], the removal of HU would also reveal a sequence dependence to loop formation.

Deletion of HU does in fact restore sequence-dependent loop formation. After deletion of HU, the repression for both sequences decreased, consistent with previous studies [5]. However, repression for constructs containing the random sequence decreased more than constructs containing the flexible sequence as seen in Figure 6A. For the flexible sequence, the looping energy was reduced by approximately 0.3 to 1.3 k$_B$T, and in the random sequence the looping energy was reduced by approximately 0.8 to 2.2 k$_B$T. The influence of HU on repression varies with operator distance, as shown in Figure 6B. This dependence on length may be due to the differential influence of HU on bending versus twisting [70], combined with the fact that at some operator distances the energy required to twist the DNA and align the operators is a larger barrier to forming the loop than bending.

The molecular mechanism leading to the masking of DNA sequence-dependent flexibility by HU is not known. In the next section we use the model to explore how a nonspecific DNA-bending protein such as HU might be able to buffer away the sequence dependence of loop formation.

**Adapting the thermodynamic model to explore the roles of sequence and DNA-bending proteins in loop formation**

To determine how the action of DNA-bending proteins can mask the intrinsic flexibility of the DNA in loop formation, we extended the thermodynamic model to include loops containing a DNA-bending protein, state 8 in Figure 7A. In this model, looping occurs either through a mechanism involving a DNA-bending protein (assisted looping), or through a mechanism independent of additional *in vivo* factors (unassisted looping), as shown in Figure 7B. By splitting the looped conformations into assisted and unassisted states, we can use this new model to predict in which regimes of parameter space DNA-bending proteins or the sequence of the loop are key factors in determining loop formation.

The terms in the weights of Figure 7A are as described for Figure 2, adding $\Delta F_{loop,u}(L)$ which is the unassisted looping energy for a given sequence and $\Delta F_{loop,a}(L)$ which is the assisted looping energy for a given sequence. Repression for this model can be expressed as,

$$\text{Repression}_{loop,2\,\text{modes}}(L) = \frac{1 + \frac{2R}{N_{NS}}(e^{-\beta\Delta\varepsilon_{rad}} + e^{-\beta\Delta\varepsilon_{rmd}}) + \frac{4R(R-1)}{N_{NS}^2}e^{-\beta(\Delta\varepsilon_{rad}+\Delta\varepsilon_{rmd})} + \frac{2R}{N_{NS}}e^{-\beta(\Delta\varepsilon_{rad}+\Delta\varepsilon_{rmd})}\left(e^{-\beta\Delta F_{loop,u}(L)} + e^{-\beta\Delta F_{loop,a}(L)}\right)}{1 + \frac{2R}{N_{NS}}e^{-\beta\Delta\varepsilon_{rad}}}. \quad (5)$$

Compared to Equation 4, incorporating assisted loop formation into the model adds another looping energy to Equation 5. The experimental results in Figure 5A show that repression for the random and flexible sequences are equivalent, leading us to postulate that the dominant mode of looping does not depend upon the sequence of the loop. Because this sequence-independent mode is dominant, it will have a lower looping energy and will be the same for both looping sequences, as depicted in Figure 7B.

Our *in vitro* results shown in Figure 4 indicate that there is sequence dependence to the intrinsic ability of DNA to form loops. The absence of any extra proteins in the *in vitro* assay shows that the unassisted looping mode has such a sequence-dependent looping energy. This is shown schematically in Figure 7B, where the unassisted looping energy for the flexible sequence is lower than the unassisted looping energy for the random sequence. We introduce the following quantities to simplify the analysis. First, the difference between the looping energies of the random and flexible sequences in the unassisted case

$$\sigma = \Delta F_{loop,u,random}(L) - \Delta F_{loop,u,flexible}(L). \quad (6)$$

Second, the difference between the looping energy for the flexible sequence in the unassisted case and the looping energy in the assisted case (which we are assuming to be the same for both sequences)

$$\delta = \Delta F_{loop,u,flexible}(L) - \Delta F_{loop,a}(L). \quad (7)$$

Redefining the energies is this way, we can derive the repression equations for the flexible and random looping sequences as

$$\text{Repression}_{loop, 2\,\text{modes, flexible}}(L) = \frac{1 + \frac{2R}{N_{NS}}(e^{-\beta\Delta\varepsilon_{rad}} + e^{-\beta\Delta\varepsilon_{rmd}}) + \frac{4R(R-1)}{(N_{NS})^2}e^{-\beta(\Delta\varepsilon_{rad}+\Delta\varepsilon_{rmd})} + \frac{2R}{N_{NS}}e^{-\beta(\Delta\varepsilon_{rad}+\Delta\varepsilon_{rmd}+\Delta F_{loop,a}(L))}(1+e^{-\beta\delta})}{1 + \frac{2R}{N_{NS}}e^{-\beta\Delta\varepsilon_{rad}}}, \quad (8)$$

and

$$\text{Repression}_{loop, 2\,\text{modes, random}}(L) = \frac{1 + \frac{2R}{N_{NS}}(e^{-\beta\Delta\varepsilon_{rad}} + e^{-\beta\Delta\varepsilon_{rmd}}) + \frac{4R(R-1)}{(N_{NS})^2}e^{-\beta(\Delta\varepsilon_{rad}+\Delta\varepsilon_{rmd})} + \frac{2R}{N_{NS}}e^{-\beta(\Delta\varepsilon_{rad}+\Delta\varepsilon_{rmd}+\Delta F_{loop,a}(L))}(1+e^{-\beta(\delta+\sigma)})}{1 + \frac{2R}{N_{NS}}e^{-\beta\Delta\varepsilon_{rad}}}. \quad (9)$$

In Figure 7C and D we examine the consequences of two modes of loop formation on the repression levels using Equations 8 and 9. In both cases the unassisted looping energy for the flexible sequence is 2 $k_BT$ lower than for the random sequence ($\sigma$ = 2 $k_BT$). This value of $\sigma$ corresponds to the *in vitro* experimental results in Figure 4C.

When including an energetically favorable, sequence-independent looping mechanism, it is possible to mask the effects of sequence flexibility. With $\delta$=0, there is a clear difference in repression between the random and flexible sequences as shown in Figure 7C. However as $\delta$ increases, corresponding to a larger energy difference between the assisted and unassisted loops, the difference in repression between the two sequences decreases and becomes immeasurable when $\delta$ is 2 $k_BT$, as shown in Figure 7D. Supplementary Figures 3 and 4 further explore how the energy level differences of the looping states dictate probabilities of looping and whether or not sequence dependence will be observed in gene regulation. From the data of Becker and coworkers [5], Supplementary Figure 4B shows that deleting HU decreased the looping energy up to 2-3 $k_BT$, an amount sufficient to hide sequence-dependent flexibility in our experimental system. This suggests that the *in vivo* results shown in Figure 6 are consistent with a mechanism in which assisted looping causes gene regulation to be independent of the looping sequence.

In Equation 9, the weight of the looping term is modulated by $(1+e^{-\beta(\delta+\sigma)})$, in which the first term is associated with the assisted looping state and the second term accounts for the extra energy needed to form the unassisted loop. Assuming that the assisted looping energy is less than or equal to the unassisted looping energy, we see that once $\delta$ is large, the value of $(1+e^{-\beta(\delta+\sigma)})$ goes to 1 regardless of the value of $\sigma$. Therefore, a large offset of the assisted and unassisted looping energies will mask sequence dependence.

### **DNA-bending protein HU assists loop formation *in vitro***

We further explore the role of HU on loop formation by using the tethered particle motion assay. As in Figure 4 we monitor the ability of the Lac repressor to loop double-stranded DNA, but here we fix the length of the loop at an operator spacing of 141.5 bp and titrate purified HU protein into our experimental setup. The operators are out of phase at this spacing, as shown in Figure 4, causing the looping probability to be low in the absence of HU. To our knowledge this is the first examination of the effect of HU on looping by the Lac repressor using an *in vitro* assay that can directly detect loop formation and breakdown. We therefore performed a number of controls to characterize the interaction of HU with the Lac repressor, which are detailed in the Supplementary Information. These controls include a demonstration that our purified HU in the absence of the Lac repressor compacts DNA tethers in a manner consistent with previous *in vitro* single-molecule studies (Supplementary Figure 5), and that the addition of HU to the TPM assay changes only the energy of loop formation by the Lac repressor, not the affinity of the Lac repressor for its operators (Supplementary Figure 6).

As shown in Figure 8A, the addition of HU dramatically increases the looping probabilities of both the flexible and random sequences. Even for these out-of-phase operators, looping probability approaches nearly 100% at high concentration of HU, demonstrating that HU can have a very large effect on the looping activity of the Lac repressor.

In the previous section, we examined the consequences of an assisted looping state. Next we extended the model to explicitly include the role of HU in loop formation. As shown in Supplementary Figure 10, in this model HU can bind in the looping region and reduce the looping energy. From previous experiments the binding affinity of HU to double-stranded DNA is approximately 480 nM [71, 72], and *E. coli* contains about 30,000 HU proteins per cell [73]. In Figure 8A, we see if the *in vitro* data is consistent with a two-state model of looping, shown with the dotted lines. In the two-state looping model, the loop can form either in the absence or presence of HU binding in the loop, as shown in Figure 8B. Using the measurement at 0 HU, the looping energies for the unassisted looping states were calculated, and from the HU titration data we calculated the looping energies for the HU bound loops. These values are listed in Figure 8C.

With these parameters, we can now predict whether HU is able to buffer away sequence dependence at the *in vivo* level of HU molecules per cell. Notice that fits to the two-state model result in the looping energy of the HU bound loops to depend on the loop sequence, 17.4 $k_BT$ for the random sequence and 16.2 $k_BT$ for the flexible sequence. We calculate repression for the two-state model using the looping energies extracted from the *in vitro* data and Equation S8 in Supplementary Information which was derived using the states and weights of Supplementary Figure 10. Supplementary Figure 11 shows that the two-state model predicts the flexible sequence will repress more than three times the random sequence at the *in vivo* HU concentration. Given that for the two-state model the sequences have different looping energies for the HU bound loop, it is not surprising that the flexible sequence would be predicted to repress more. The predicted sequence dependence of repression is inconsistent with *in vivo* results in Figure 5 which show that both looping sequences repress to the same extent. Therefore although the two-state model fits well to the *in vitro* data shown in Figure 8A, it does not correctly predict *in vivo* regulation.

In order to obtain sequence-independent repression *in vivo*, both the flexible and random sequences seem to require equivalent looping energies for the HU bound state. Forcing the looping energies of the HU bound state to be the same for both the flexible and random sequences results in a best-fit looping energy of 16.9 $k_BT$. In Supplementary Figure 12, we see that this value is not in good agreement with the *in vitro* data for either the flexible or random sequence. The failure of the two-state looping model to predict sequence-independent repression *in vivo* suggests that this simplest model of a single HU binding in the loop is not sufficient to describe both the *in vitro* and *in vivo* results.

To explore the possibility of a model in which more than one HU can bind in the loop, we implement a three-state looping model, as depicted in Figure 8B. The details of this model can be found in the Supplementary Information. The solid lines in Figure 8A show a fit to the *in vitro* data using the three-state model while constraining the looping energy for the doubly HU bound loop to be equal for both looping sequences. Adding the extra looped state with this constraint ensures that repression will be sequence-independent at the high concentration of HU found *in vivo*, as shown in Supplementary Figure 11. We observe that the three-state model with this constraint remains in good agreement with the *in vitro* data. In the three-state model, loop formation with 0 and 1 HU in the loop is sequence-dependent, and loop formation with 2 HU bound in the loop is sequence-independent. In this way, HU in the loop gradually reduces the sequence dependence of loop formation. The *in vitro* results demonstrate that loop formation is strongly influenced by the presence of HU, and support a model of looping formation in which multiple HU proteins bind in the loop.

**CONCLUSION AND OUTLOOK**

The ability of transcription factors to bind to non-adjacent sites on the genome and to act at a distance requires the intervening DNA to bend, twist, or loop into configurations which bring these DNA-bound gene regulatory factors into proximity. DNA sequence through its influence on DNA mechanics is potentially a key determinant of gene regulation in such contexts, since all of these regulatory mechanisms involve mechanical distortions of the DNA [1, 17]. The goal of this paper has been to examine the role of sequence-dependent free energy of DNA deformation as a potentially biologically relevant control parameter of gene regulation in the bacterial setting. In this work, we quantify the relative roles of sequence and DNA-bending proteins in determining the mechanical properties of DNA *in vivo*. We focus on flexible sequences containing TA steps, with anisotropic bendability, as opposed to previously measured poly(A) tracts with isotropic bendability [37, 41]. Our strategy was to utilize quantitative models of gene regulation that predict the role of sequence-dependent mechanical properties in gene regulation, and to test these predictions through both *in vitro* and *in vivo* experimental measurements.

Our simple model predicted that loop-mediated gene regulation would be very responsive to changes in the mechanical properties of the loop. To test this prediction we compared repression for two different looping sequences that were shown to have different *in vitro* flexibilities as shown in Figure 2C and 4C. However, the expected increase in repression for the flexible looping sequence was not observed *in vivo*. Upon further model development and parallel *in vivo* and *in vitro* experiments, we revealed that the ability of the DNA-bending protein HU to assist in loop formation through a sequence-independent mechanism masks the intrinsic sequence dependence of loop formation. HU and the functionally related protein IHF have been reported to modulate the overall degree of looping by the Lac repressor, both *in vivo* and *in vitro* [11, 40, 52]. Our results build on these previous examinations to characterize the interplay between the intrinsic flexibility of DNA set by sequence and assisted bending by DNA-structuring proteins in the determination of loop-mediated gene regulation. We have shown here that intrinsic, sequence-dependent flexibility may not in fact be a parameter that sets gene expression levels in bacteria. These results demonstrate that looping assisted by DNA-bending proteins *in vivo* can complicate the picture of DNA mechanics derived from *in vitro* experiments.

We quantified the role of HU in loop formation in single-molecule *in vitro* measurements. Using these measurements, we were able to extend our theoretical models to incorporate HU binding to the loop. The *in vitro* data was consistent with HU binding in the loop and lowering the free energy change of looping. However, the simplest model of a single HU protein abolishing sequence dependence of loop formation was not supported. Instead, the data support a model in which multiple HU proteins can bind to the looping region. In this model, each HU lowers the energy of loop formation and partially diminishes the influence of the looping sequence on loop formation. As the concentration of HU increases, looping gradually becomes both more probable and less dependent on sequence, as shown in Supplementary Figure 13. A short DNA loop with multiple HU proteins bound is feasible. Two HU proteins in a Lac repressor-mediated loop has been demonstrated in structural calculations and this conformation is proposed to enable new looping geometries [61]. Future work involving the Forster resonance energy transfer may shed light on how the number of HU proteins bound in a looping region influences the dynamics and geometry of loop formation [74]. HU is at a high concentration *in vivo*, estimated to be bound every 550 bp during exponential phase [68], and only occupies between of 9 and 30 base pairs when bound [68]. The ubiquity, small binding footprint, and relative nonspecific binding of HU to double-stranded DNA suggest the *lac* operon is not a special case, and that HU may be able to buffer away sequence-dependent mechanical properties in other regulatory contexts throughout the genome.

There are many DNA-bending and chromosomal architectural proteins known in *E. coli* besides HU including IHF, Fis, and H-NS [44, 68].  The differences in repression in the absence of HU between the two sequences are still less than predicted in Figure 3D given the sequence-dependent energy differences observed in Figure 4, suggesting that DNA-bending proteins other than HU may also contribute to loop formation *in vivo* and may still be partially masking the sequence-dependent looping energy in the ΔHU strains.  In addition, HU and other nucleoid proteins have a myriad of direct and indirect effects on global gene regulation [44], making it difficult to isolate the specific role any of these factors have on loop formation *in vivo*.  It is as yet unclear how multiple DNA-bending proteins potentially work together to influence mechanical deformation of DNA.  Further studies that sequentially remove each of these other proteins *in vivo* or combine them together *in vitro* will be needed to determine which are redundant in masking sequence dependence in repression.  Moreover, our models for the interaction of HU within Lac repressor-mediated loops is not detailed enough to account for the particular role of HU in DNA loop geometry.  Further investigation into the interaction between HU and sequence in terms of twist flexibility and loop orientation is warranted.

As there are some reports that HU binds to specific sequences or specific DNA structures [11], it is interesting that the flexibilities for both the flexible and random sequences were equal in the presence of HU.  Our results suggest the flexible sequence did not form any structures or contain any sequences which attracted HU more than the random sequence.  Although the sequences used here were the most flexible and inflexible sequences identified by one study in the context of nucleosome and minicircle formation [24], it is possible that other sequences not tested here will lead to altered repression even in the presence of HU.  A study which specifically selects for mechanical properties on the basis of repressor-mediated loop formation may identify sequences with an increased or decreased propensity for looping even in the presence of HU.  It seems likely that some operons may contain specialized sequences that modulate looping probabilities.  Given that DNA deformation is ubiquitous in many cellular processes, including gene regulation, it will be interesting to see in which situations sequence-dependent mechanical properties are relevant and in which cases DNA-structuring proteins play a dominant mechanistic role.  Such a quantitative understanding of DNA mechanics and how it is encoded in the genome will be a critical part of understanding and predicting gene regulation at base pair resolution.

**ACKNOWLEDGEMENTS**

We would like to dedicate this paper to the memory of Jon Widom, who was a fundamental creative force in the conception and implementation of this experiment. We are grateful to Bob Schleif and Tom Kuhlman for helpful discussions and technical assistance, to Reid Johnson for the HU-expressing strain used in this work, to Sankar Adhya for the HU purification protocol, and to both Reid Johnson and Sankar Adhya for advice and help on the HU expression and purification.  HU was purified in collaboration with Jost Vielmetter and Angela Ho at the Caltech Protein Expression Center. This work was supported by the National Institutes of Health [awards No. DP1 OD000217 (HGG and RP), No. R01 GM085286, No. R01 GM085286-01S (HGG, JQB, and RP), and 1 U54 Ca143869 (Northwestern PSOC Center)], NSF graduate fellowship (SJ) and La Fondation Pierre Gilles de Gennes (RP).

Author contributions: J.Q.B., H.G.G., S.J., and R.P. designed research; J.Q.B., H.G.G., and S.J. performed research; and J.Q.B., H.G.G., S.J., and R.P. analyzed data and wrote the paper.

**FIGURE LEGENDS**

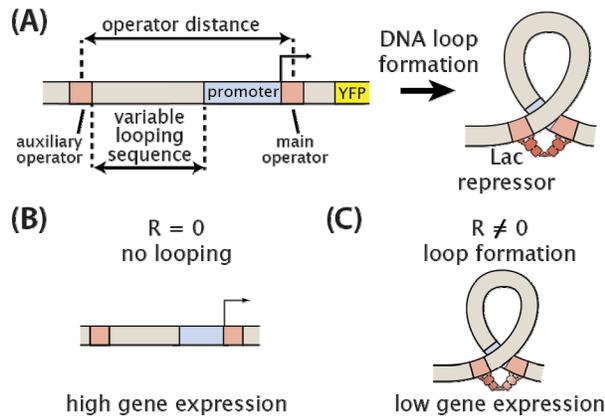

**Figure 1: Loop-mediated repression of gene expression.** (A) We have created a suite of synthetic YFP expression constructs, in which the promoter expressing YFP is under negative control by the *E. coli* Lac repressor. There are two binding sites for the repressor in the vicinity of the promoter: a main operator located at +11 relative to the transcription start site and an auxiliary operator located upstream from the promoter at a variable distance. Lac repressor can bind to both operators simultaneously, looping the intervening DNA. There are four commonly used *lac* operators, each with a different affinity for Lac repressor. Here we use Oid as the auxiliary operator and O2 as the main operator. The variable region of the loop is derived either from a synthetic, random E8 sequence, or a putatively very flexible TA sequence [30]. The operator distance is defined from the center of each operator. (B) Placing constructs into host strains that do not express Lac repressor results in high expression of the YFP reporter gene. R is the number of repressors per cell. (C) Placing constructs into host strains that express Lac repressor results in loop formation which reduces expression of the reporter gene.

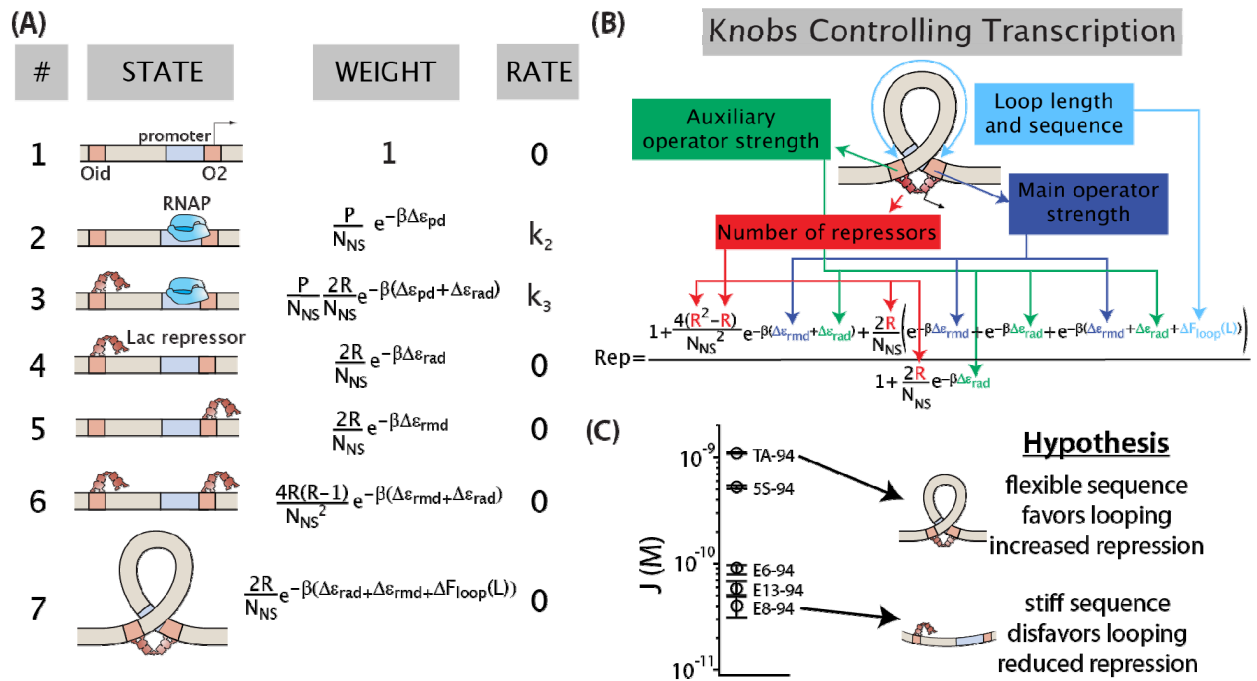

**Figure 2: A quantitative model of gene expression predicts how cellular parameters tune the level of repression.** (A) The states of the DNA looping constructs, their associated weights, and the rates of transcription from each state used in the thermodynamic model. Refer to text for a description of the different variables. (B) From the statistical mechanical model, an expression for the experimentally measurable quantity repression was derived, Equation 4. This equation quantifies how each "knob" of the system (operator binding energies, number of repressors per cell, loop sequence and length) modulates the reduction of gene expression. We use this expression to predict how repression (Rep) will be influenced by the flexibility of the looping sequence. (C) The two looping sequences used in this work represent the extremes of flexibility observed by [24], as measured by the J-factor of cyclization (a higher J-factor indicates increased flexibility). Here we test the hypothesis that a more flexible looping sequence will increase repression *in vivo*.

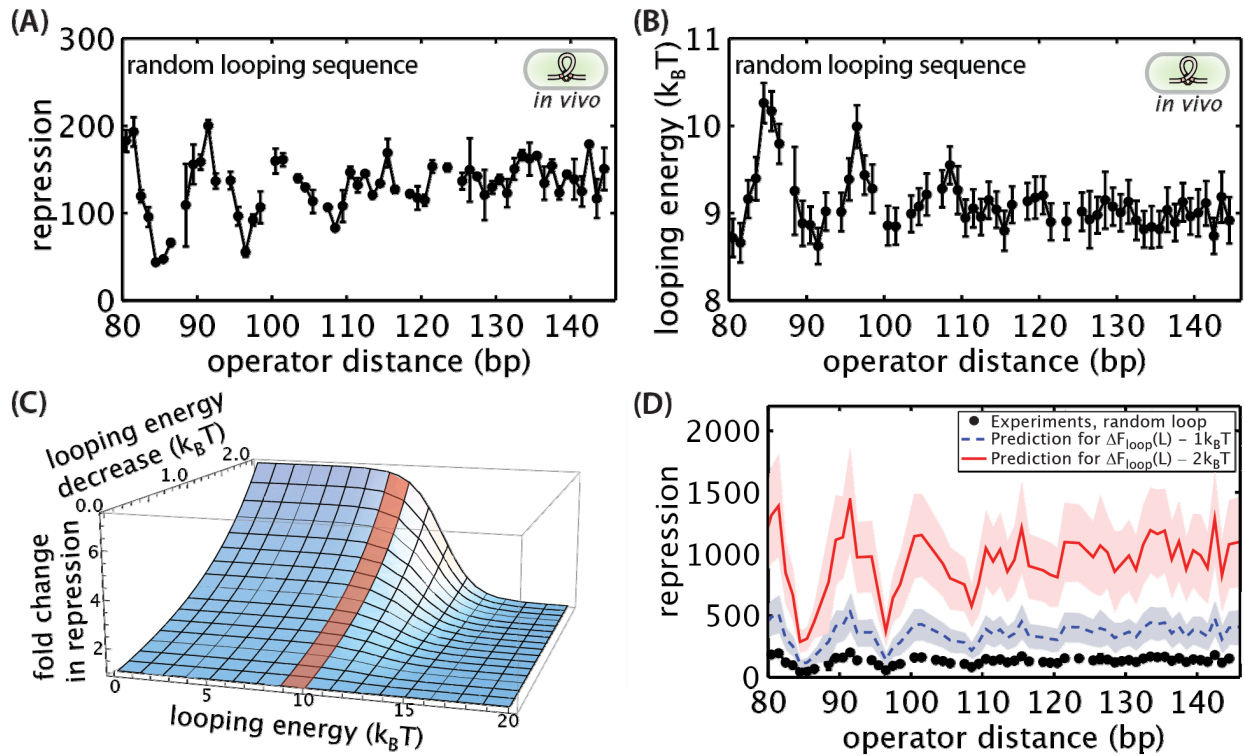

**Figure 3: Prediction of repression for a flexible looping sequence.** (A) Repression for operator distances between 80.5 and 145.5 bp are shown for the random loop sequence for constructs containing the operators O2 and Oid as main and auxiliary operators, respectively. (B) Using Equation S4, the looping free energy was extracted from the repression data for each operator distance. The data in (A) and (B) were previously reported in [48]. (C) Using the results from (B) as a starting point, repression was found to be sensitive to decreases in the looping energy over the range of looping energies for the random sequence (red shaded region). (D) Predicted repression using Equation 4 when the looping energy is decreased by 1 or 2 $k_BT$ for operator distances between 80.5 and 145.5. The previously reported results shown in (A) were used as a basis for the predictions. Shaded regions in (D) represent standard error of the prediction. Error bars correspond to the standard error.

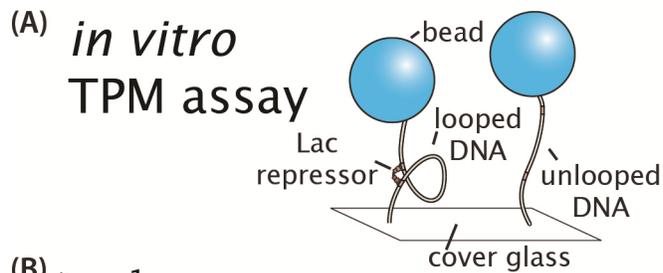
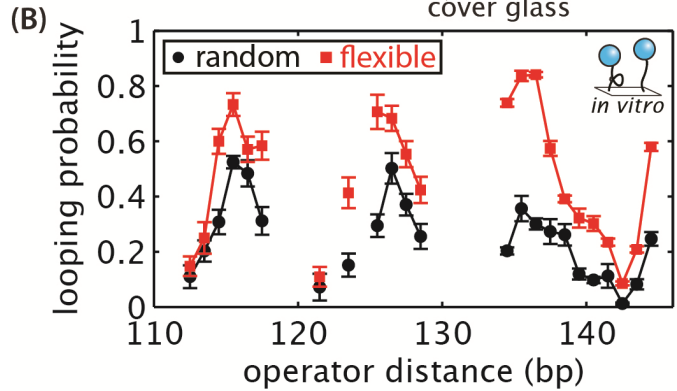
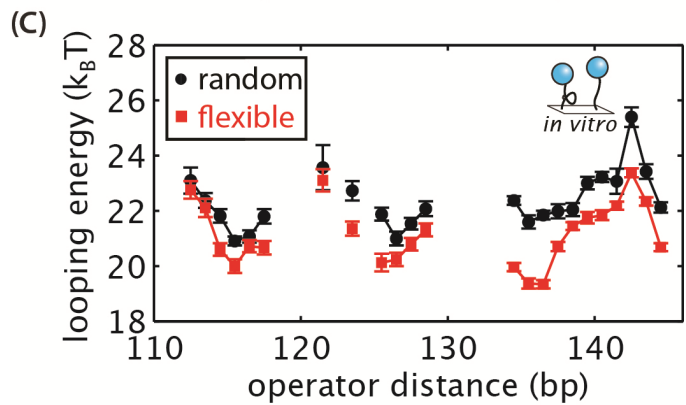

**Figure 4: Calculation of the looping energy for the random and flexible DNA sequences from *in vitro* measurements of looping probability.** (A) The tethered particle motion (TPM) assay was used to quantify the *in vitro* mechanical properties of the random and flexible looping sequences. (B) Results previously reported in [51] show the probability of looping for each sequence as determined by measuring the change in the length of the DNA tether over time in the presence of Lac repressor. (C) Using the previously reported results in (B), the looping energies of each sequence were calculated. The flexible sequence lowers the looping energy by 1-3 $k_B T$ at many loop lengths. Error bars correspond to standard errors for looping probabilities and bootstrapped errors for looping energies [51].

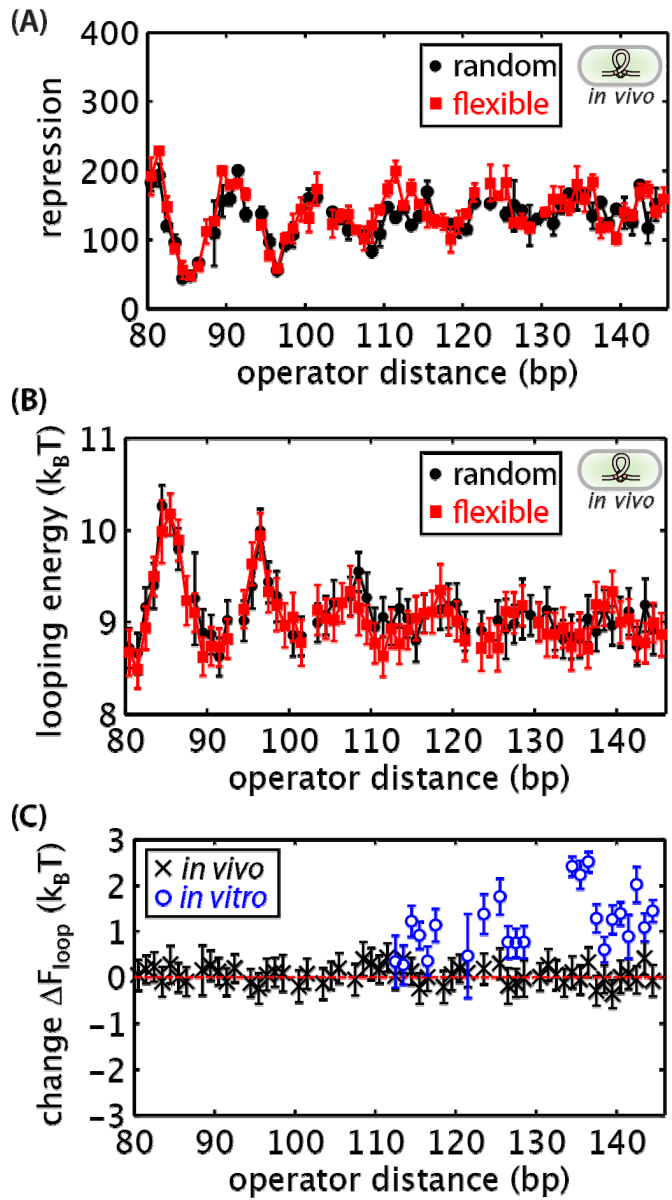

**Figure 5: *In vivo* loop-mediated repression.** *In vivo* assay for loop-mediated repression using a fluorescent gene reporter was used to compare repression (A) and looping energies (B) for the random and flexible looping sequences. (C) Plotting the difference in the observed looping energy between the random and flexible sequences at each loop length emphasizes that both sequences have similar propensities to form loops *in vivo*. In contrast, the *in vitro* looping free energy of the random sequence can be more than 2 $k_BT$ greater than the flexible sequence. The dashed red line corresponds to no difference between the looping energies. Error bars correspond to standard errors.

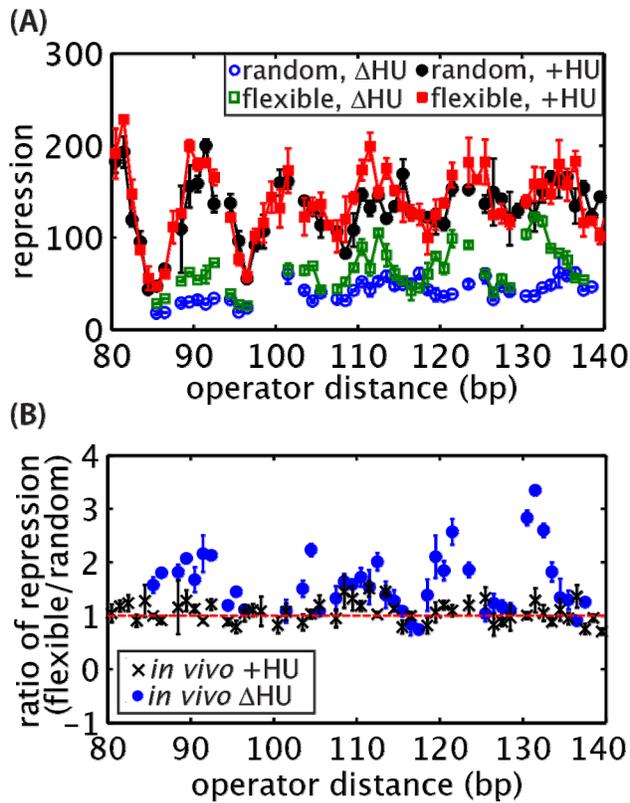

**Figure 6: Deletion of HU restores sequence dependence of loop-mediated repression.** (A) Repression in the absence ("ΔHU") versus presence ("+HU") of the nonspecific DNA-bending protein HU. Consistent with previous reports [5], deletion of HU decreases repression at all lengths. However, we show here that the presence of HU also masks a sequence dependence to looping that is only detectable when HU is deleted. (B) Comparison of the ratio in repression observed for the random and flexible sequences with and without HU. Error bars are standard errors.

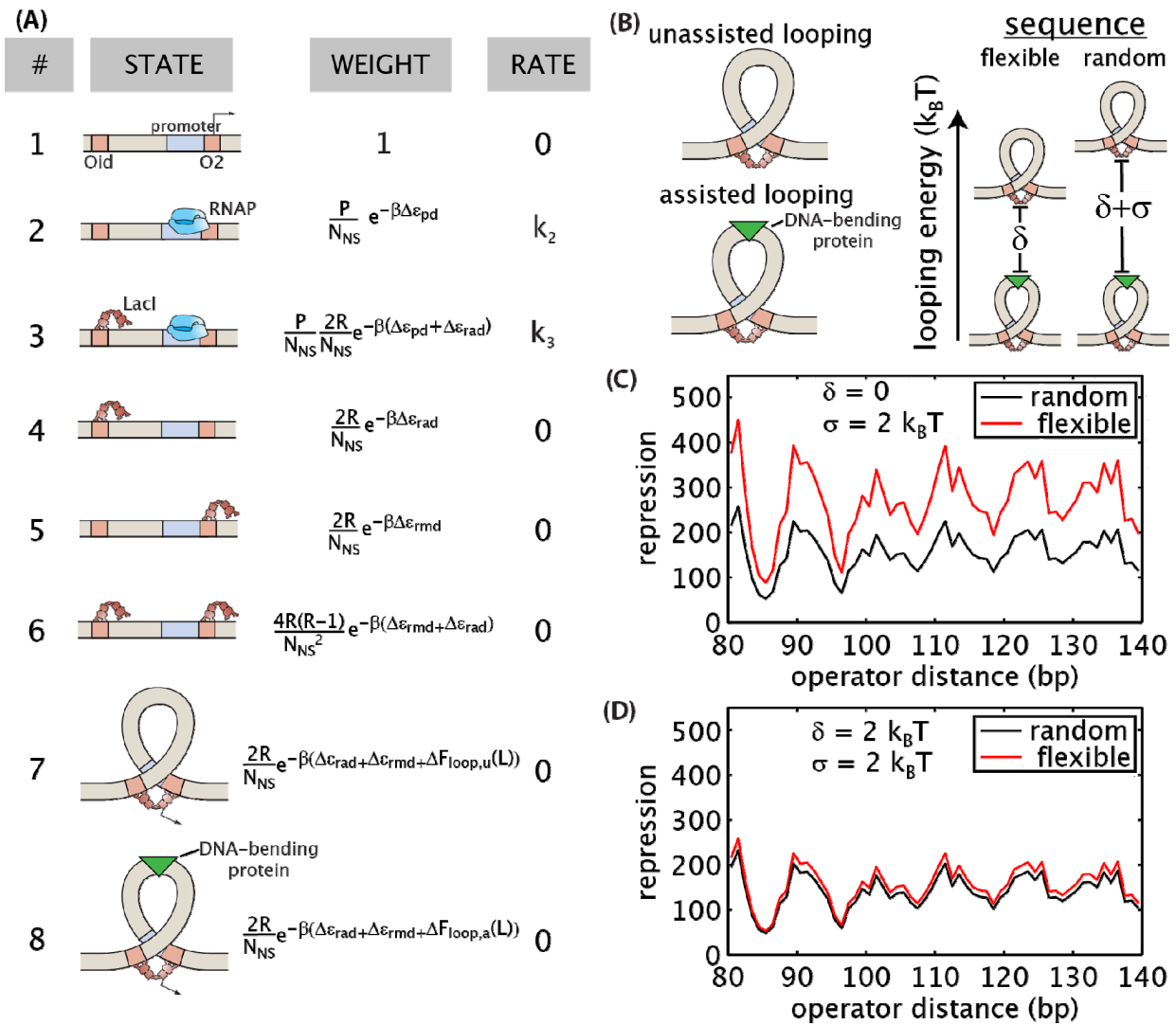

**Figure 7: Modeling sequence-dependent buffering by assisted loop formation.** (A) A model that incorporates two modes of loop formation adds a new state of assisted looping, state 8, to the model shown in Figure 2A. (B) Given that repression in wild type cells is not dependent on sequence as shown in Figure 5, we assume the assisted looping energies for both the random and flexible sequences are equal. The unassisted looping energy for the flexible sequence is δ $k_BT$ greater than the assisted looping energy, and the unassisted looping energy for the random sequence is (δ+σ) $k_BT$ greater than the assisted looping energy. (C-D) Using experimental data from Figure 5A as a starting point, when the difference between the unassisted looping energies for the flexible and random sequences, σ, is 2 $k_BT$, the model predicts how repression will change as δ is increased. Calculations use Equations 8 and 9. (C) For δ = 0, the flexible sequence represses more than the random sequence. (D) For δ = 2 $k_BT$, loops containing flexible sequences of DNA repress similarly to loops containing random sequences.

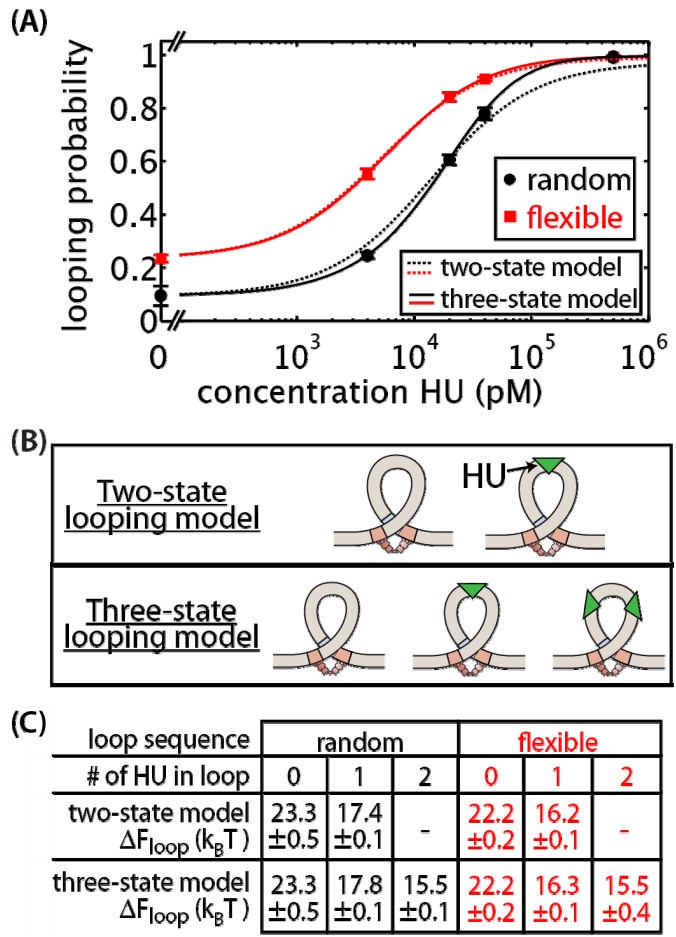

**Figure 8: Effect of HU on *in vitro* loop formation.** (A) Probability of looping as a function HU concentration for both the flexible (red squares) and random (black circles) looping sequences. Loop formation in the presence of 100 pM Lac repressor was measured using a tethered particle motion *in vitro* assay. Dotted lines show the fit of a two-state looping model, in which the loop can form with either 0 or 1 HU molecules in the looping region as shown in the top of (B). Solid lines show the fit of a three-state looping model, in which the loop can form with 0, 1, or 2 HU molecules in the loop as shown in the bottom of (B). For the flexible looping sequence, the fit to the two- and three-state models almost completely overlap. (C) Looping energies ± standard error fit to the *in vitro* data using either the two-state or three-state looping model. See Supplementary Information for further details. Error bars are standard errors. Broken axis in (A) is used to show the looping probabilities at 0 HU.